\documentclass{article}
\usepackage[utf8]{inputenc}
\usepackage{authblk}
\usepackage{amsmath}
\usepackage{amssymb}
\usepackage{hyperref}

\title{The LoDoPaB-CT Dataset \\ \small{A Benchmark Dataset for Low-Dose CT Reconstruction Methods}}
\author[1]{Johannes Leuschner \thanks{\href{mailto:jleuschn@uni-bremen.de}{jleuschn@uni-bremen.de}}}
\author[1]{Maximilian Schmidt \thanks{\href{mailto:maximilian.schmidt@uni-bremen.de}{maximilian.schmidt@uni-bremen.de}}}
\author[1]{Daniel Otero Baguer \thanks{\href{mailto:otero@uni-bremen.de}{otero@uni-bremen.de}}}
\author[1]{\\Peter Maaß \thanks{\href{mailto:pmaass@uni-bremen.de}{pmaass@uni-bremen.de}}}
\affil[1]{University of Bremen, Center for Industrial Mathematics}
\date{October 2019}

\usepackage[minbibnames=3, maxbibnames=3, giveninits=true]{biblatex}
\usepackage{graphicx}
\usepackage{geometry}
\usepackage{siunitx}
\usepackage{amsthm}
\usepackage{algorithm}
\usepackage{algorithmic}
\usepackage{subcaption}
\usepackage{xcolor}
\usepackage{booktabs}
\usepackage{tikz}

\theoremstyle{remark}
\newtheorem*{remark}{Remark}

\begin{document}


\maketitle
\begin{abstract}
    {Deep Learning approaches for solving Inverse Problems in imaging have become very effective and are demonstrated to be quite competitive in the field. Comparing these approaches is a challenging task since they highly rely on the data and the setup that is used for training. We provide a public dataset of computed tomography images and simulated low-dose measurements suitable for training this kind of methods. With the LoDoPaB-CT Dataset we aim to create a benchmark that allows for a fair comparison. It contains over \num{40000} scan slices from around \num{800} patients selected from the LIDC/IDRI Database. In this paper we describe how we processed the original slices and how we simulated the measurements. We also include first baseline results.}
\end{abstract}

\section{Introduction}
\label{sec:introduction}
{Tomographic image reconstruction is an extensively studied field.
Traditionally, analytical methods, like filtered back-projection (FBP), or iterative reconstruction (IR) techniques are used for this task. Recently, image reconstruction approaches involving Deep Learning (DL) have been developed and demonstrated to be very competitive \cite{adler2017iterative_nn, chen2017convnet_ct, jin2017cnn_imaging, li2018nett, lunz2018adversarial, shan2019nn_vs_vendor_ct, wang2018image_reconstruction, yang2018gan_ct}.

One popular imaging modality is Computed Tomography (CT). As high doses of applied radiation are potentially harmful to the subjects, better reconstruction methods for low-dose scans are highly desirable. Different learned methods have been successfully applied to this task \cite{adler2017iterative_nn, wang2018image_reconstruction}. However, comparing these approaches is a challenging task since they highly rely on the data and the setup that is used for training. The main goal of this work is to provide a standard dataset that can be used to train and benchmark learned low-dose CT reconstruction methods.}

{To this end we introduce the Low-Dose Parallel Beam (LoDoPaB)-CT, which uses the public LIDC/IDRI Database \cite{armato2011lidc_idri} of human chest CT reconstructions. We consider these (in the form of 2D images) to be the so called ground truths. Paired samples, created by simulating low-dose CT measurements, constitute the most complete training data and could be used for all kinds of learning. In particular, methods trained only on prior samples, i.e.\ CT images, and methods relying on samples of the joint distribution of CT images and the measurements, can make use of (part of) the dataset. In total, the dataset features more than \num{40000} sample pairs.}

The dataset is published on \href{https://zenodo.org}{zenodo.org} and accessible via DOI \href{https://doi.org/10.5281/zenodo.3384092}{\texttt{10.5281/zenodo.3384092}}.\footnote{\url{https://doi.org/10.5281/zenodo.3384092}} It is subject to the Open Data Commons Attribution License (ODC-By) v1.0.\footnote{\url{https://opendatacommons.org/licenses/by/1.0/}}
Each dataset part is stored in multiple HDF5 files \cite{hdf5}, which can be processed with simple tools in many programming languages.
Nevertheless, the python library DIV$\alpha\ell$\footnote{\url{https://github.com/jleuschn/dival}. Library for testing and comparing DL based methods for inverse problems.} might be useful for accessing the data easily.

\section{Related Work}
\label{sec:related_work}
Machine learning based approaches benefit strongly from the availability of comprehensive datasets. In the last years a wide variety of different CT data has been published, covering different body parts and scan scenarios.

For the training of reconstruction models the projection data is crucial, but is rarely provided. Recently, Low Dose CT Image and Projection Data (LDCT-and-Projection-data) \cite{mccollough2020ldctpd} was published by investigators from the Mayo Clinic, which includes measured normal-dose projection data in the new open DICOM-CT-PD format of \num{299} patients. The AAPM Low Dose CT Grand Challenge data \cite{mcCollough2016dataset_grand_ct_challenge} is another dataset that includes (simulated) sinograms, featuring \num{30} different patients. The Finish Inverse Problems Society (FIPS) provides complete datasets of a walnut \cite{haemaelaeinen2015dataset_fips_walnut} and a lotus root \cite{bubba2016dataset_fips_lotus} aimed at sparse data tomography. Recently, \textsc{Der Sarkissian} et al.\ \cite{sarkissian2019dataset_walnuts} published cone-beam CT projection data and reconstructions of \num{42} walnuts. Their dataset is directly aimed at the training and comparison of machine learning methods. In magnetic resonance imaging, fastMRI \cite{zbontar2018dataset_fastmri} with \num{1600} scans of humans knees is another prominent example.

Other CT datasets focus on the detection and segmentation of special structures like lesions in the reconstructions for the development of computer-aided diagnostic (CAD) methods. Therefore, they do not include the sinograms. FUMPE \cite{masoudi2018dataset_fumpe} contains CT angiography images of $35$ subjects for the detection of pulmonary embolisms. KiTS2019 \cite{heller2019dataset_kits19} is build around the segmentation of kidney tumor in CT images. The Japanese Society of Radiology Technology (JSRT) database \cite{shiraishi2000dataset_jsrt_database} and the National Lung Screening Trial (NLST) in cooperation with the CT Image Library (CTIL) \cite{clark2007dataset_ctil, cody2010dataset_nlst} each contain scans of the lung. 

The LIDC/IDRI Database \cite{armato2011lidc_idri, armato2011lidc_idri_data} consist of \num{1018} helical thoracic CT scans (cf.\ section \ref{sec:lidc_idri}). It forms the basis for our LoDoPaB-CT Dataset. We simulate the missing projection data  to make the dataset usable for the training of reconstruction algorithms. The whole process is described in section \ref{sec:simulation_setup}.

\section{CT Imaging}
\label{sec:ct_imaging}
Introduced in the 1970s, computed tomography has become one of the most valuable technologies in medical imaging. A mathematical description of CT was developed by \textsc{Radon} \cite{radon1986radon_trafo} over 50 years earlier. This section briefly introduces the mathematical foundation and a model for the stochastic noise.

\begin{figure}[bt]
    \centering
    \begin{subfigure}[b]{0.323\textwidth}
        \includegraphics[width=\textwidth]{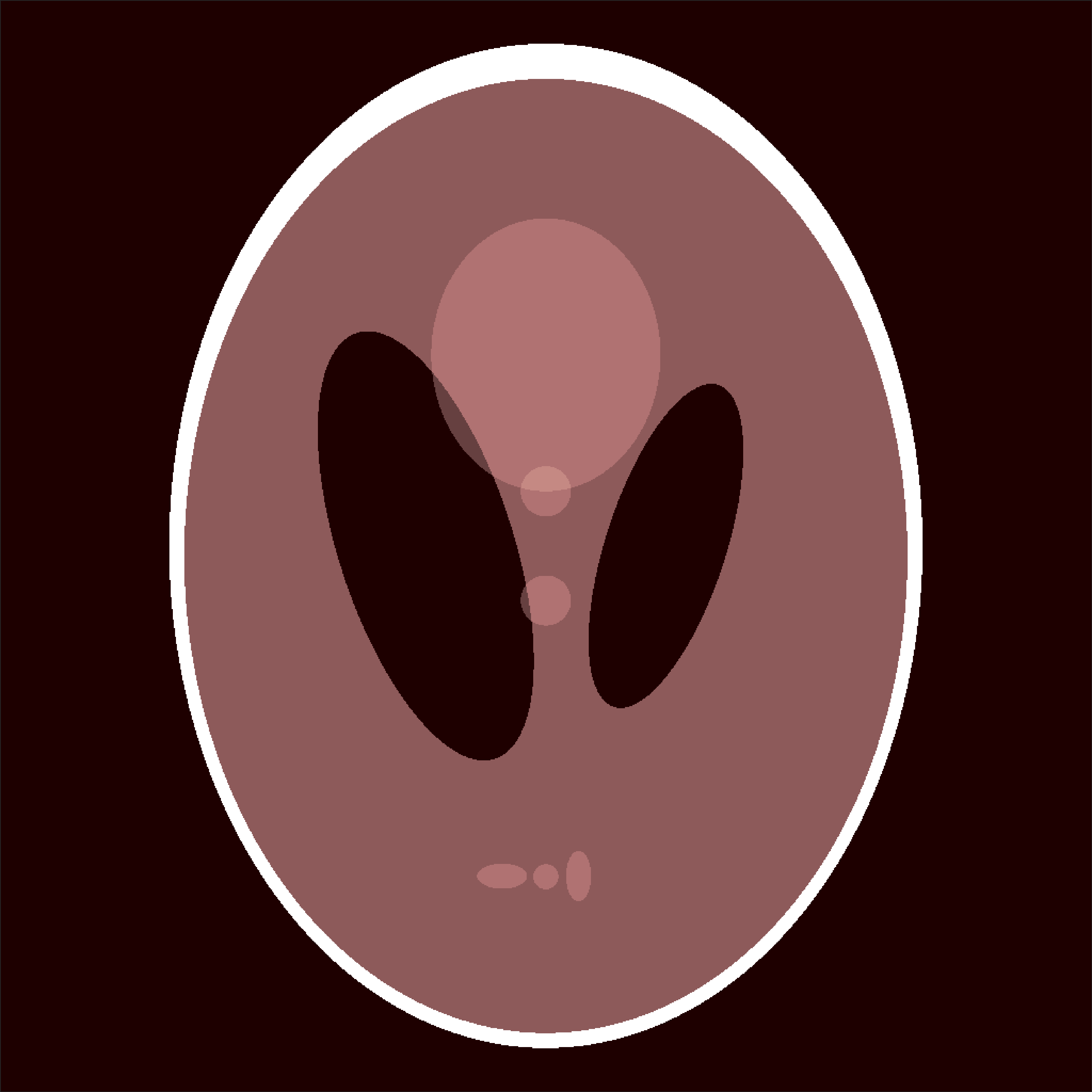}\vspace{0.98em}
    \end{subfigure}
    ~ 
    \begin{subfigure}[b]{0.46\textwidth}
        \includegraphics[width=\textwidth]{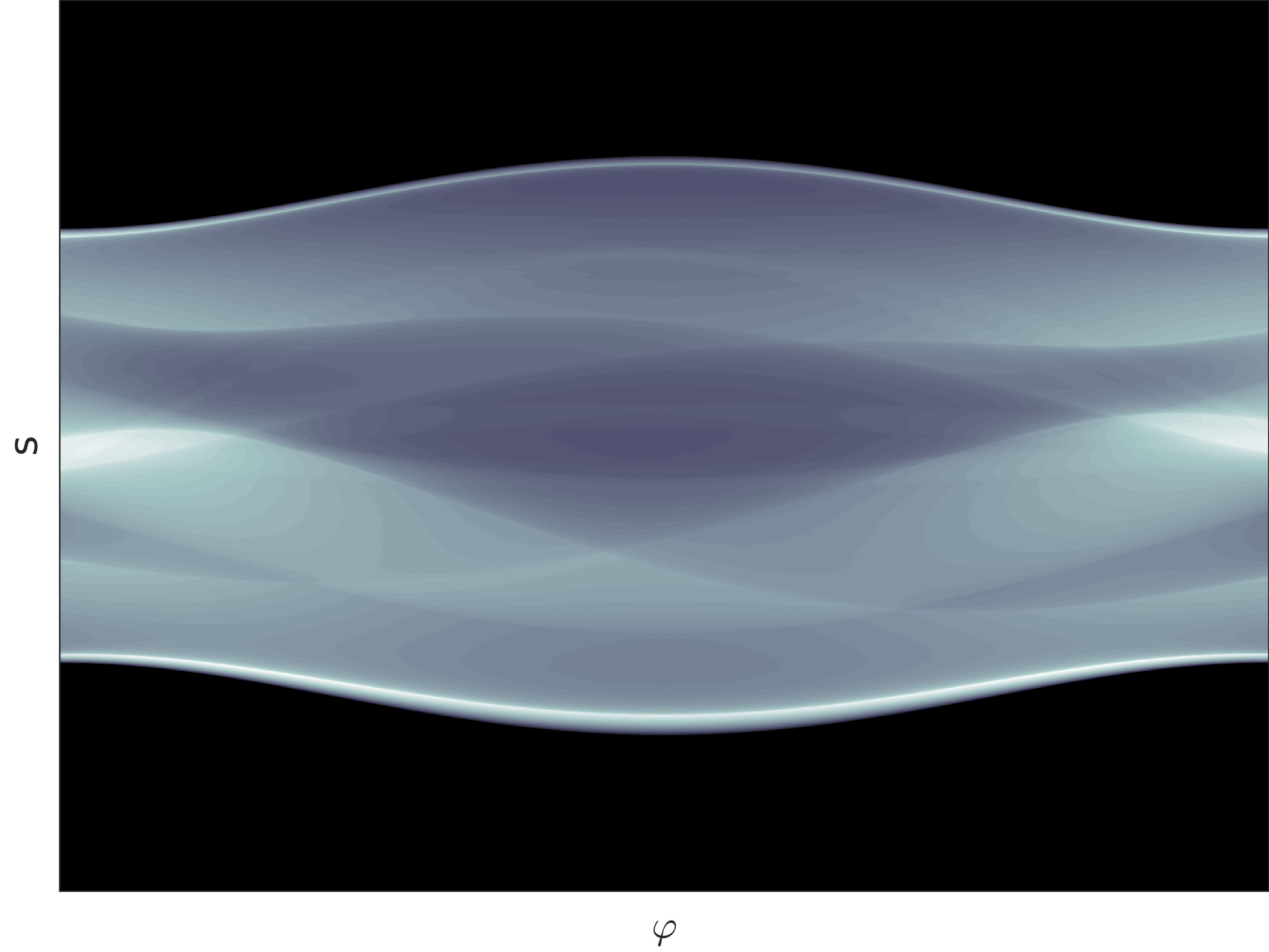}
    \end{subfigure}
    \caption{The Shepp-Logan phantom (left) and its corresponding sinogram (right).}
    \label{fig:shepp-logan_sinogram}
\end{figure}

\begin{figure}[H]
    \centering
    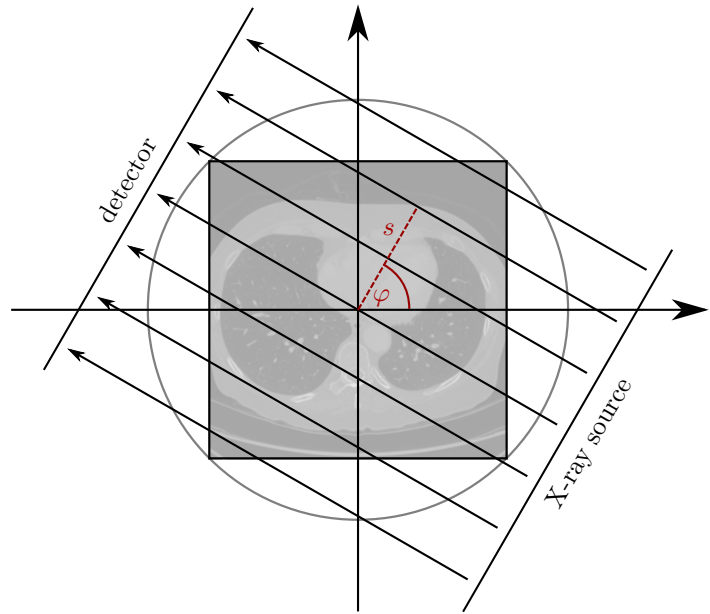
    \caption{Parallel beam geometry}
    \label{fig:parallel_beam}
\end{figure}

\subsection{Mathematical Foundation}
The task in CT belongs to the class of \textit{inverse problems}
\begin{align}
    \mathcal{A} x + \varepsilon = y^{\delta}.
    \label{eq:inverse_problem}
\end{align}
\noindent Given noisy measurements $y^{\delta}$, so called \textit{sinograms}, the goal is to calculate an accurate reconstruction $\tilde{x}$ of the interior distribution $x$ of a body $\Omega$, with respect to some quality measure. The noise $\varepsilon$ has a level of $\delta$.

For CT we use the 2D Radon transform as forward operator $\mathcal{A}$ to model the attenuation of the X-ray when passing trough body $\Omega$ with interior distribution $x$. We can parameterize the path of an X-ray beam by the distance from the origin $s \in \mathbb{R}$ and angle $\varphi \in [0,\pi]$ (cf.\ Figure \ref{fig:parallel_beam})
\begin{align*}
    L_{s,\varphi}(t) = s \omega\left(\varphi\right) + t \omega^\perp\left(\varphi\right), \quad \omega\left(\varphi\right):= \begin{bmatrix}\cos(\varphi) \\ \sin(\varphi) \end{bmatrix}, \quad \omega^\perp\left(\varphi\right) :=  \begin{bmatrix}-\sin(\varphi) \\ \phantom{-}\cos(\varphi) \end{bmatrix}.
\end{align*}
The Radon transform then calculates the integral along the line for parameters $s$ and $\varphi$
\begin{align}
    \mathcal{A} x(s,\varphi) = \int_{\mathbb{R}} x\left(L_{s,\varphi}(t) \right) \, \mathrm{d}t.
\end{align}
According to Beer-Lambert's law, the result is the logarithm of the ratio between the intensity $I_0$ at the X-ray source and $I_1$ at the detector
\begin{align}
    \mathcal{A} x(s,\varphi) = -\ln \left(\frac{I_1\left(s, \varphi\right)}{I_0\left(s, \varphi\right)}\right) = y\left(s, \varphi\right).
\end{align}
Calculating the transform for all pairs $(s, \varphi)$ results in a sinogram. An example for the famous Shepp-Logan phantom is shown in Figure \ref{fig:shepp-logan_sinogram}.

To get a reconstruction $\tilde{x}$ from the sinogram, we have to invert the forward model. Since the Radon transform is linear and compact, the inverse problem is \textit{ill-posed} in the sense of \textsc{Nashed} \cite{nashed1987ill_posed, natterer2001math_tomography}.
In practice, we can only measure for a finite number of distances $s$ and angles $\varphi$. In the discrete case the problem is ill-conditioned.

\subsection{Photon Statistics and Noise}
Noise in CT can be classified into \textit{quantum noise} and \textit{detector noise}. The whole process of photon generation, attenuation and detection can be modelled by a Poisson distribution,
while the detector noise is often assumed to be Gaussian. The number of measured photons $N_{1}$ at the detector is therefore given by
\begin{align}
    N_{1}(s, \varphi) = n_{1}(s, \varphi) + \varepsilon, \quad n_{1}(s, \varphi) \sim \operatorname{Poisson}\left(N_{0} \exp\left(-\mathcal{A}x(s, \varphi) \right) \right), \quad \varepsilon \sim \mathcal{N}\left(0, \sigma^2 \right).
    \label{eq:noise_model}
\end{align}
For the expected number of photons follows
\begin{align}
    \mathbb{E}\left[N_{1}(s, \varphi)\right] = \mathbb{E}\left[n_{1}(s, \varphi)\right] + \mathbb{E}\left[\varepsilon\right] = N_0 \exp\left(- \mathcal{A}x(s, \varphi) \right),
    \label{eq:expected_photon_count}
\end{align}
which is again Beer-Lambert's law. Besides the stochastic fluctuations, artifacts can occur due to material properties, physical effects or movement of the patient during the scan (cf.\ \textsc{Buzug} \cite{buzug2008computed_tomography}).

\section{LIDC/IDRI Database and Data Selection}
\label{sec:lidc_idri}
The Lung Image Database Consortium (LIDC) and Image Database Resource Initiative (IDRI) published the LIDC/IDRI Database \cite{armato2011lidc_idri} to support the development of CAD methods for the detection of lung nodules. The dataset consists of \num{1018} helical thoracic CT scans of \num{1010} individuals. Seven academic centers and eight medical imaging companies collaborated for the creation of the database. As a result, the data is heterogeneous with respect to the technical parameters and scanner models.

\begin{figure}[t]
    \centering
    \includegraphics[width=0.31\textwidth]{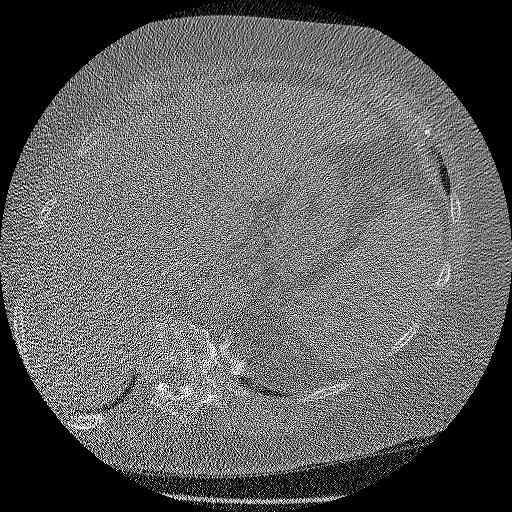}
    \hfill
    \includegraphics[width=0.31\textwidth]{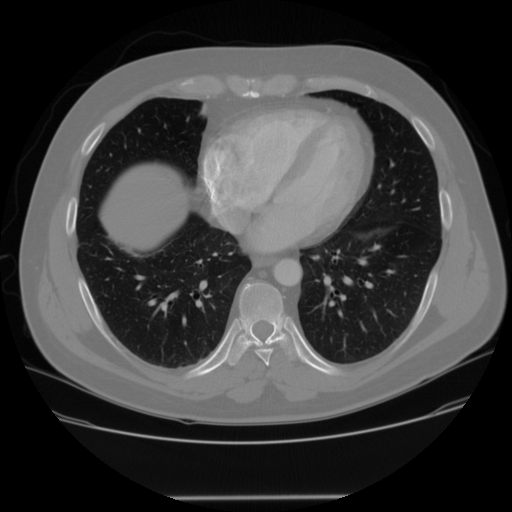}
    \hfill
    \includegraphics[width=0.31\textwidth]{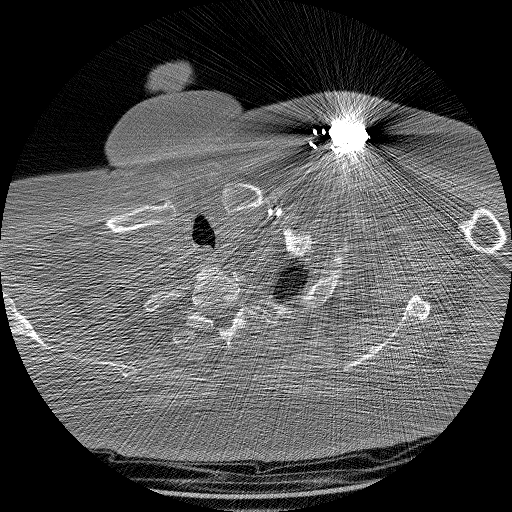}
    \caption{Scans from the LIDC/IDRI Database with poor quality, good quality and an artifact. The shown HU window is $[-1024, 1023]$.}
    \label{fig:lidc_idri_examples}
\end{figure}

Both standard-dose and lower-dose scans are part of the dataset. Tube peak voltages range from $\SI{120}{kV}$ to $\SI{140}{kV}$ and tube current from $\SI{40}{mA}$ to $\SI{627}{mA}$ with a mean of $\SI{222.1}{mA}$. Labels for the lung nodules were create by a group of \num{12} radiologists in a two-phase process. The image reconstruction was performed with different convolution kernels, depending on the manufacturer of the scanner. Figure \ref{fig:lidc_idri_examples}
shows examples of the provided reconstructions. The LIDC/IDRI Database is freely available from The Cancer Imaging Archive (TCIA) \cite{clark2013tcia}. It is published under the Creative Commons Attribution 3.0 Unported  License\footnote{\url{https://creativecommons.org/licenses/by/3.0/}}.

The LoDoPaB-CT Dataset builds on the LIDC/IDRI scans. Our dataset is intended for the evaluation of reconstruction methods in a low-dose setting. Therefore, we simulate the projection data, which is not included in the LIDC/IDRI dataset (cf.\ section \ref{sec:simulation_setup}). To guarantee a fair comparison with good ground truths, patients whose scans are too noisy were removed in a manual selection process.  Additional scans were excluded due to their geometric properties. In the end, \num{812} patients remain in the LoDoPaB-CT Dataset.

The dataset is split into four parts: three parts for training, validation and testing, respectively, and a ``challenge'' part that is kept private for the time being.\footnote{If you are interested in participating in our challenge-like comparison, please contact us via \href{mailto:jleuschn@uni-bremen.de}{jleuschn@uni-bremen.de}.}
Each part contains scans from a distinct set of patients as we want to study the case of learned reconstructors being applied to patients that are not known from training.
The training set features scans from \num{632} patients, while the other parts contain scans from \num{60} patients each.
Every scan contains multiple slices (2D images) for different $z$-positions, of which only a subset is included.
The amount of extracted slices depends on the slice thickness obtained from the metadata.\footnote{As slices with small distances are similar, they may not provide much additional information while increasing the chances to overfit.
With this reasoning we attempted to only include slices that have a minimum distance of \SI{7.5}{mm} to each other, which we stated in the previous version of this paper (\href{https://arxiv.org/abs/1910.01113v1}{arXiv:1910.01113v1}). However, we have to correct ourselves, since we mistakenly assumed the slice files to be ordered by their $z$-positions when using the filename as sort key (and also assumed the distances to be equal to the slice thickness). The actual distances of the extracted slices are larger than \SI{5.0}{mm} for $>\SI{45}{\%}$ and larger than \SI{2.5}{mm} for $>\SI{75}{\%}$ of the slices.}
In total, the dataset contains \num{35820} training images, \num{3522} validation images, \num{3553} test images and \num{3678} challenge images.

\begin{remark}
 We propose to use our default dataset split, as it allows for fair comparison with other methods that use the same split.
 However, users are free to remix or resplit the dataset parts.
 For this purpose randomized patient IDs are provided, i.e., the same random ID is given for all slices obtained from one patient.
 Thus, when creating custom splits it can be regulated whether---and to what extent---the same patients are contained in different parts.
\end{remark}

\section{Simulation Setup}
\label{sec:simulation_setup}
For the simulated CT measurements the following setup was chosen:
\begin{itemize}
 \item parallel beam geometry (cf.\ Figure \ref{fig:parallel_beam})
 \item \num{1000} projection angles
 \item \num{513} projection beams (lowest number meeting Nyquist-criterion)
 \item Poisson-distributed noise matching a mean photon count of \num{4096} per detector pixel before attenuation.
\end{itemize}

The computation steps that were applied to the images from the LIDC/IDRI Database are listed in algorithm \ref{alg:simulation} in more detail and will be discussed in the next sections.
For technical reference, the script used for simulation is available online.\footnote{\url{https://github.com/jleuschn/lodopab_tech_ref}}

\subsection{Image preprocessing}

First of all, each image is cropped to the central rectangle of $\SI{362}{px} \times \SI{362}{px}$.
This is done because most of the images contain (approximately) circle-shaped reconstructions with a diameter of $\SI{512}{px}$ (cf.\ Figure \ref{fig:lidc_idri_examples}).
After the crop, the image only contains pixels that lie inside this circle, which avoids value jumps occuring at the border of the circle.
While this yields natural ground truth images, we need to point out that the cropped images in general do not show the full subject but some interior part. Hence, it is unlikely for methods trained with this dataset to perform well on full-subject measurements.

For some scan series, the circle is subject to a geometric transformation either shrinking or expanding the circle in some directions.
In particular, for few scan series the circle is shrinked such that it is smaller than the cropped rectangle.
We exclude these series, i.e.\ those with patient IDs 0004, 0032, 0102, 0116, 0120, 0289, 0368, 0418, 0541, 0798, 0926, 0972 and 1000, from our dataset, which allows to crop all included images consistently to $\SI{362}{px} \times \SI{362}{px}$.

The integer Hounsfield unit (HU) values obtained from the DICOM files are dequantized by adding uniform noise from the interval $[0, 1)$.
Next, the linear attenuations $\mu$ are computed from the dequantized HU values using the definition of the HU,
\begin{align}
 \mathrm{HU} = 1000\, \frac{\mu-\mu_\mathrm{water}}{\mu_\mathrm{water}-\mu_\mathrm{air}} \qquad \Leftrightarrow \qquad \mu = \mathrm{HU}\, \frac{\mu_\mathrm{water}-\mu_\mathrm{air}}{1000} + \mu_\mathrm{water},
 \label{eqn:hu2mu}
\end{align}
where we use the linear attenuation coefficients
\begin{align}
 \mu_\mathrm{water} &= \SI{20}{\per m},\\
 \mu_\mathrm{air} &= \SI{0.02}{\per m},
\end{align}
which approximately correspond to an X-ray energy of \SI{60}{keV} \cite{hubbell1995mass_attenuation}.
Finally, the $\mu$-values are normalized into $[0, 1]$ by dividing by
\begin{align}
 \mu_\mathrm{max} = 3071\, \frac{\mu_\mathrm{water}-\mu_\mathrm{air}}{1000} + \mu_\mathrm{water} = \SI{81.35858}{\per m},
\end{align}
which corresponds to the largest HU value that can be represented with the standard 12-bit encoding, i.e. $(2^{12}-1-1024)\,\si{HU} = \SI{3071}{HU}$, followed by clipping of all values into $[0, 1]$,
\begin{align}
 \hat\mu = \mathrm{clip}(\mu / \mu_\mathrm{max}, [0, 1]) = \begin{cases}
 0 &, \mu / \mu_\mathrm{max} \leq 0\\
 \mu / \mu_\mathrm{max} &, 0 < \mu / \mu_\mathrm{max} \leq 1\\
 1 &, 1 < \mu / \mu_\mathrm{max}
 \end{cases}.
 \label{eqn:mu_normed}
\end{align}
The equations \eqref{eqn:hu2mu} and \eqref{eqn:mu_normed} are applied pixel-wise to the images.

\begin{algorithm}[t]
\caption{Simulation}
\begin{algorithmic}[1]
 \setlength{\baselineskip}{1.5\baselineskip}
 \vspace{0.1em}\item[] $\mu_\mathrm{air} = \num{0.02}, \mu_\mathrm{water} = \num{20}, \mu_\mathrm{max} = \num{81.35858}$
 \vspace{0.1em}\item[] \textit{Preprocessing:}\\
 \STATE image\_HU = dcm.pixel\_array[75:$-75$, 75:$-75$] $*$ dcm.RescaleSlope $+$ dcm.RescaleIntercept
 \STATE image\_HU += dequantization\_noise $\sim \mathcal{U}(0, 1)$
 \STATE image\_mu = image\_HU $*$ ($\mu_\mathrm{water}$ $-$ $\mu_\mathrm{air}$) $/$ 1000 $+$ $\mu_\mathrm{water}$
 \STATE \textbf{image} = clip(image\_mu $/$ $\mu_\mathrm{max}$, min=0, max=1)
 \vspace{0.1em}\item[] \textit{Simulation:}\\
 \STATE image\_upscaled = bilinear\_interp(image\_mu, 1000$\times$1000)
 \STATE proj\_data = ray\_trafo(image\_upscaled, rect=$[\num{-0.13},\num{0.13}]^2$, angles=1000, detector\_pixels=513)
 \STATE photons = exp($-$proj\_data) $*$ \num{4096}
 \STATE noisy\_photons $\sim$ max(\num{0.1}, Pois(photons))
 \STATE \textbf{measurement} = $-$log(noisy\_photons $/$ \num{4096}) $/$ $\mu_\mathrm{max}$
\end{algorithmic}
\label{alg:simulation}
\end{algorithm}

\subsection{Simulation of CT measurements}

To simulate the measurements based on the virtual ground truth images, the main step is to apply the forward operator, which is the ray transform for CT.
For this task we utilize the Operator Discretization Library (ODL) \cite{odl} with the \texttt{'astra\_cpu'} backend \cite{astra_toolbox}.\footnote{We choose \texttt{'astra\_cpu'} over the usually favored \texttt{'astra\_cuda'} because of small inaccuracies observed in the sinograms when using \texttt{'astra\_cuda'}, specifically at angles $0$, $\frac{\pi}{2}$ and $\pi$ and detector positions $-1/\sqrt{2}\,\frac{l}{2}$ and $1/\sqrt{2}\,\frac{l}{2}$ with $l$ being the length of the detector. The used version is \texttt{astra-toolbox==1.8.3} on Python 3.6. The tested CUDA version is 9.1 combined with \texttt{cudatoolkit==8.0}.}

In order to avoid ``committing the inverse crime'' \cite{wirgin2004inverse_crime}, which, in our scenario, would be to use the same discrete model both for simulation and reconstruction, we use a higher resolution for the simulation.
Otherwise, a good performance of reconstructors for the specific resolution of this dataset ($362\times 362$) could also stem from the properties of the specific discretized problem, rather than from good inversion of the analytical model.
We use bilinear interpolation for the upscaling of the virtual ground truth from $362\times 362$ to $1000\times 1000$.

The non-normalized, upscaled image is projected by the ray transform.
By Beer-Lambert's law, the intensity quotient $I_1/I_0$ then is $\exp(-y)$, where $y$ is the result of the projection.
In order to compute the photon count per detector pixel, this quotient is multiplied with the number of photons per detector pixel before attenuation, which is chosen to be \num{4096}.
This photon count is used as the mean (and variance) of the Poisson distribution from which the measured photon count is sampled.
In order to avoid photon starvation, photon counts that are zero are replaced by a photon count of \num{0.1}.
Hereby strictly positive values are ensured, which is a prerequisite for the log-transform in the next step (see e.g.\ \cite{wang2017hybrid_pre_post_log}).
The log-transform of the noisy intensity quotient is calculated and divided by $\mu_\mathrm{max}$ to match the normalized ground truth images.

\begin{remark}
 Although the linear model obtained by the log-transform is easier to study, in some cases pre-log models are more accurate. See \textsc{Fu} et al.\ \cite{fu2017pre_log_post_log} for a detailed comparison.
 For applying a pre-log method, the stored observation data $\hat y$ must be back-transformed by $\exp(-\mu_\mathrm{max} \cdot \hat y)$.
 In this case the ground truth images also should be multiplied with $\mu_\mathrm{max}$ to create physically consistent data pairs.
\end{remark}
\begin{remark}
 Note that the minimum photon count of \num{0.1} can be adapted subsequently.
 This is most easily done by filtering out the highest observation values and replacing them with $-\log(\epsilon / 4096) / \mu_\mathrm{max}$, where $\epsilon$ is the minimum photon count.
\end{remark}

\section{Evaluation Criteria}
\label{sec:evaluation}
The baseline models in section \ref{sec:baseline} are evaluated against two standard image quality metrics often used in CT applications \cite{joemai2017ssim_in_ct, adler2017}.
Since ground truth data is provided in the data set, we restrict ourselves to so called \textit{full-reference} methods. Namely these are the \textit{peak signal-to-noise ratio} and the \textit{structural similarity} by \textsc{Wang} et al.\ \cite{wang2004ssim}. While the first metric calculates pixel-wise intensity comparisons between ground truth and reconstruction, structural similarity captures structural distortions.

\subsection{Peak Signal-to-Noise Ratio}
The peak signal-to-noise ratio (PSNR) expresses the ratio between the maximum possible image intensity and the distorting noise
\begin{align}
    \operatorname{PSNR}\left(\tilde{x}, x\right) = 10 \log_{10} \left( \frac{\max\left(x \right)^2}{\operatorname{MSE}\left(\tilde{x}, x \right)} \right).
    \label{eq:psnr}
\end{align}
Here $x$ is the ground truth image and $\tilde{x}$ the reconstruction. Higher PSNR values are an indication for a better reconstruction. Since the reference value of $3071$ HU is far from the most common values, we choose $\max\left(x \right)$ to be the difference between highest and lowest entry in $x$ and not the maximum possible intensity. Otherwise, the results would often be too optimistic. 

\subsection{Structural Similarity}
Based on assumptions about the human visual perception, the structural similarity (SSIM) compares the overall image structure of ground truth and reconstruction. It is computed through a sliding window at $M$ locations
\begin{align}
    \operatorname{SSIM}\left(\tilde{x}, x\right) = \frac{1}{M} \sum_{j=1}^{M} \frac{\left(2 \tilde{\mu}_j \mu_j + C_1 \right) \left(2 \Sigma_j + C_2 \right) }{\left( \tilde{\mu}_j^2 + \mu_j^2 + C_1 \right) \left(\tilde{\sigma}_j^2 + \sigma_j^2 + C_2\right)},
    \label{eq:ssim}
\end{align}
where $\tilde{\mu}_j$ and $\mu_j$ are the average pixel intensities, $\tilde{\sigma}_j$ and $\sigma_j$ the variances and $\Sigma_j$ the covariance of $\tilde{x}$ and $x$ at the $j$-th local window. Constants $C_1 = (K_1 L)^2$ and $C_2 = (K_2 L)^2$ stabilize the division. Following \textsc{Wang} et al.\ \cite{wang2004ssim} we choose $K_1=0.01$ and $K_2=0.03$. The window size is $7 \times 7$ and $L=\max(x)$.

\section{Baseline Reconstructions}
\label{sec:baseline}
\begin{figure}
    \centering
    \input{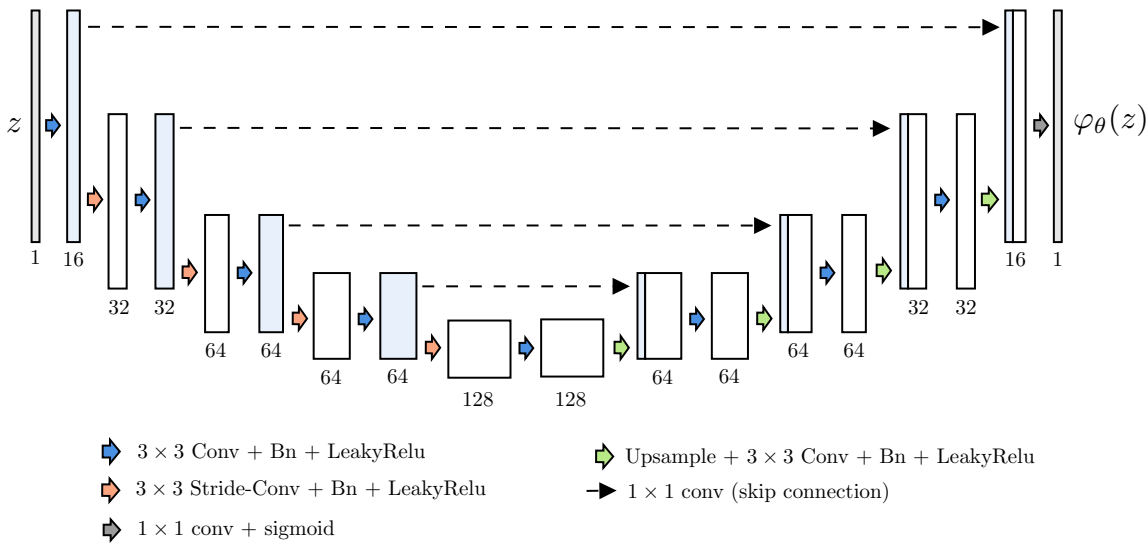}
    \caption{U-Net architecture. Numbers of channels are denoted below the layers. Skip connections have $4$ channels. Leaky ReLU activations with $a=\num{0.2}$ are used.}
    \label{fig:unet_architecture}
\end{figure}

{We provide reference reconstructions as well as quantitative results for the standard filtered back-projection (FBP) and for a learned post-processing method, which was trained using the proposed dataset. FBP is a widely used analytical reconstruction technique (cf.\ \textsc{Buzug} \cite{buzug2008computed_tomography} for an introduction). In this case we used the Ram-Lak filter without frequency scaling.

If the measurements are noisy, FBP reconstructions tend to include streaking artifacts. A typical approach to overcome this problem is to apply some kind of post-processing such as denoising. Recent works \cite{chen2017convnet_ct, jin2017cnn_imaging, yang2018gan_ct} have successfully used convolutional neural networks, such as the U-Net \cite{ronneberger2015unet}. The idea is to train a neural network to create clean reconstructions out of the noisy FBP results.

In our implementation we used a U-Net-like architecture which is shown in Figure \ref{fig:unet_architecture} and is similar to the one used in \cite{ulyanov2018dip}. There, the authors show that such an architecture is a good image prior, i.e., its output is biased towards natural images, for example, noise-free images. Training was performed by minimizing the mean squared error loss with the Adam algorithm \cite{kingma2014adam} for $20$ epochs with batch size $64$ and learning rate starting at \num{0.01} which is decayed by \num{0.1} after every $5$ epochs.
The model with the highest mean PSNR on the validation set is selected from the models reached during training.}

\begin{table}[htb]
    \centering
    \begin{tabular}{lcccccc}
         & \multicolumn{2}{c}{training set} & \multicolumn{2}{c}{validation set} & \multicolumn{2}{c}{test set}\\
         & PSNR (\si{dB}) & SSIM & PSNR (\si{dB}) & SSIM & PSNR (\si{dB}) & SSIM \\\toprule
        FBP & \num{23.99} & \num{0.408} & \num{23.94} & \num{0.407} & \num{24.43} & \num{0.426} \\\midrule
        U-Net & \num{34.48} & \num{0.838} & \num{35.00} & \num{0.858} & \num{34.24} & \num{0.820} \\\bottomrule
    \end{tabular}
    \caption{Baseline performance. Values are the mean over all samples.}
    \label{tab:baseline}
\end{table}
\begin{figure}[htb]
    \centering
    \includegraphics{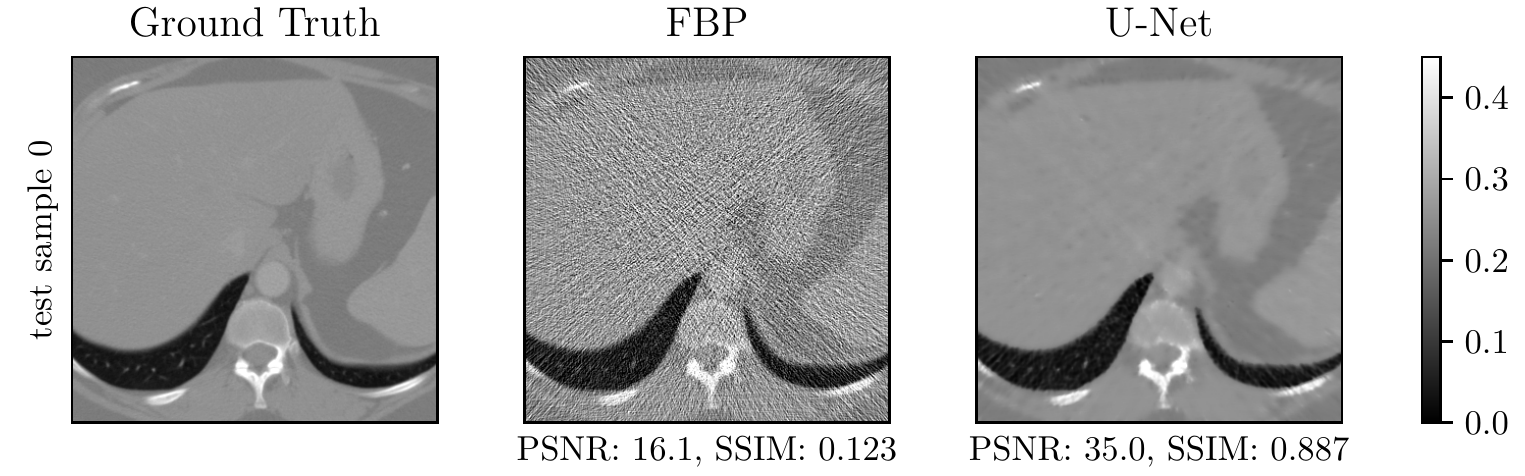}
    \includegraphics{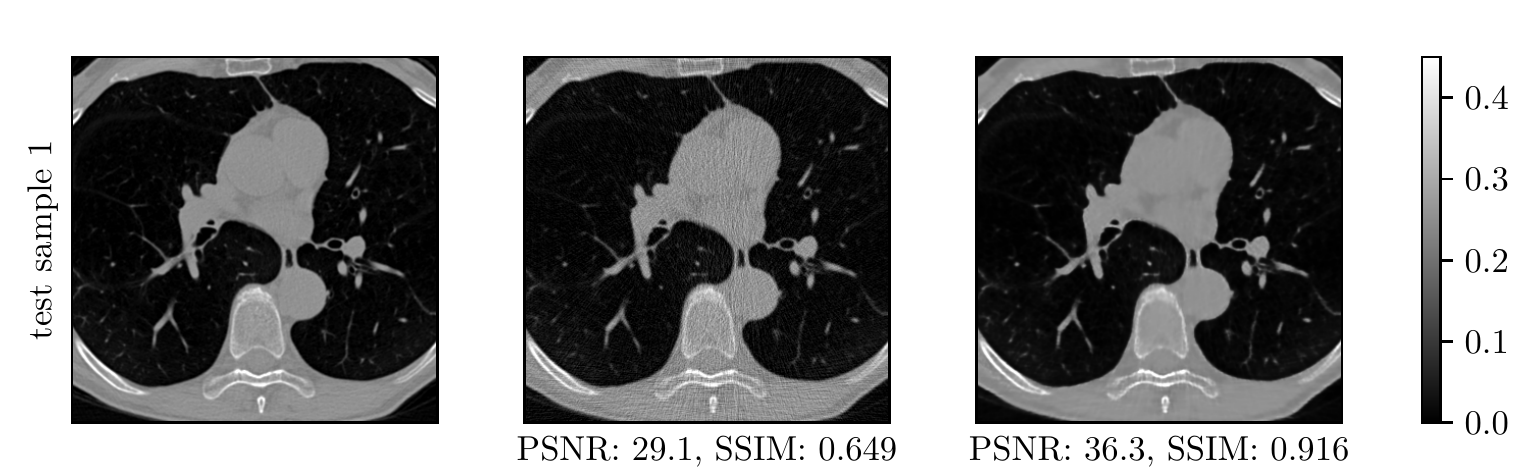}
    \includegraphics{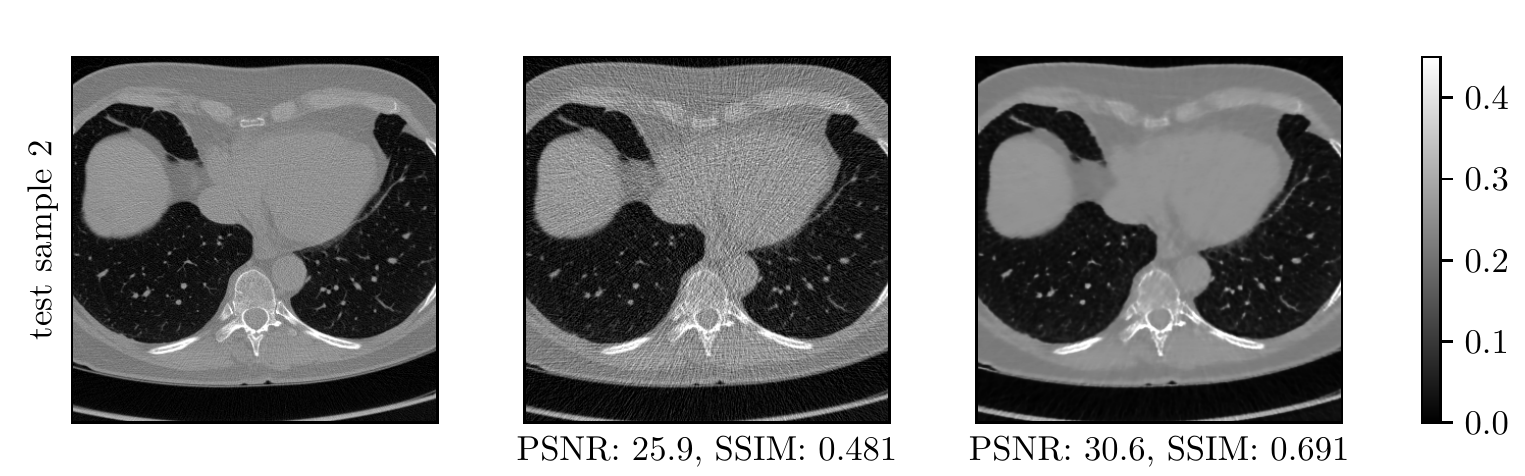}
    \includegraphics{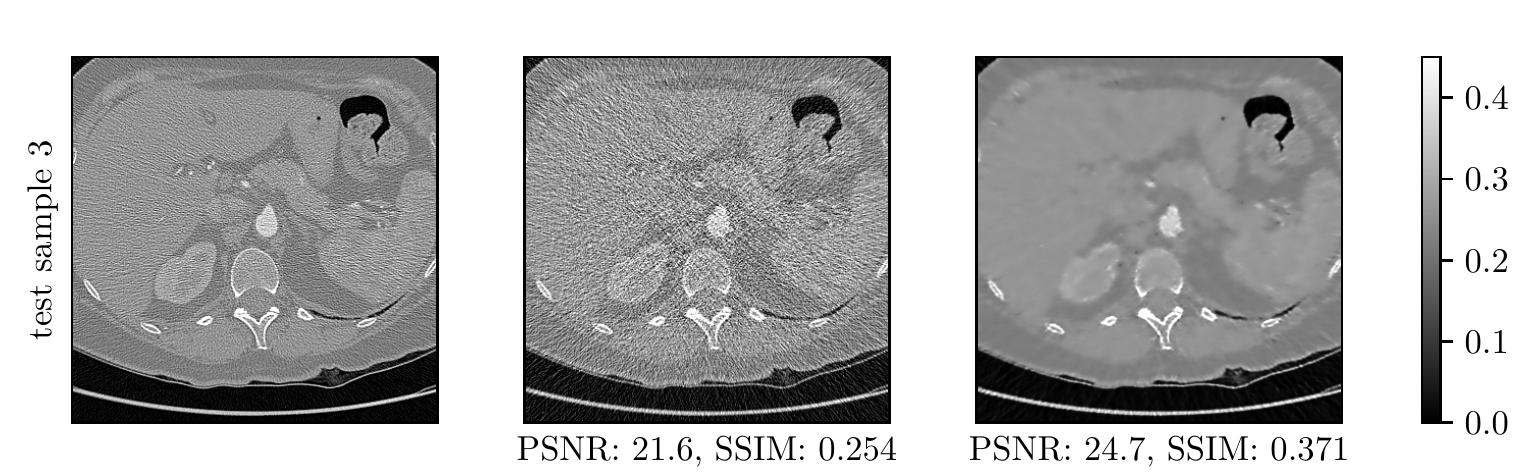}
    \caption{Baseline reconstructions. The window $[0, \num{0.45}]$ corresponds to a HU range of $\approx[-1001, 831]$.}
    \label{fig:baseline_reconstructions}
\end{figure}

In Table \ref{tab:baseline}, the performance of the baseline methods on the training, validation and test set is denoted.
Figure \ref{fig:baseline_reconstructions} shows the reconstructions for the first samples from the test set.
While the post-processing leads to major improvements in terms of the evaluated metrics, slight corruptions, such as missing details, can be observed in some of the reconstructions.
Thus, in addition to the standard metrics, which are objective and easy to evaluate, also the value of reconstructions for e.g.\ diagnosis by a medical professional should be considered.

\section{Conclusion}
\label{sec:conclusion}
{A large dataset, suitable for training deep learning approaches for low-dose CT image reconstruction, was created and made publicly available via \href{https://doi.org/10.5281/zenodo.3384092}{zenodo.org}. Human chest reconstructions from the LIDC/IDRI database were processed and filtered in order to be used as ground truths. The corresponding projection data was simulated in a low-intensity setting by applying Poisson noise. We demonstrate its usability by means of two baseline reconstruction methods, FBP (not learned) and post-processing (learned). These were applied and evaluated using standard metrics. The results confirm the potential of learned reconstruction methods over analytical ones, especially in imaging applications such as CT. A ``challenge'' part of the data is kept private for the time being in order to be used in a challenge-like comparison.}

\section{Acknowledgements}
\label{sec:acknowledgements}
We thank Simon Arridge, Ozan Öktem and Carola-Bibiane Schönlieb for fruitful discussion about the procedure, and Felix Lucka and Jonas Adler for their ideas and helpful feedback on the simulation setup.

\newpage
\clearpage
\printbibliography

\end{document}